**Localization of quantum objects in an expanding universe and cosmologically induced classicality**

C. L. Herzenberg

Introduction

We live in an expanding universe that is finite in its observed spatial extent and has existed for only a finite duration of time. But standard quantum theory is set in a static universe that is infinitely extended in both space and time. Generally, we are comfortable in using standard quantum mechanics to describe objects in our real universe, based on the implicit assumption that the very small effects of expansion in our immediate neighborhood of the universe, and the long ago/far away edge effects of our universe have no significant role to play in affecting quantum objects in the here and now. But is that assumption really valid, that the expansion and limits of our universe would have no appreciable effect on quantum objects?

In standard quantum theory, an unconstrained freely moving object is described by a plane wave that extends uniformly throughout all of infinite space-time. This is a completely unlocalized wave function having equal probability density everywhere, so that a free object in standard quantum theory would have equal probability to be found anywhere in the universe. But in the real world that we live in, macroscopic objects are always observed in spatially well-localized states, behaving as classical objects. This is quite in contrast to micro-objects that can usually be found in energy eigenstates, behaving quantum mechanically. So, the question arises, might the localization of free macroscopic objects have its origin in the finite but expanding character of our space-time?

The answer may very well be 'yes', and there are several studies by Paul Davies and myself that propose and support the idea that the classical behavior of macroscopic objects can be cosmologically caused. Since I have only a short time to speak about this, I will not go into Davies approach and his results, as I would like to concentrate on the studies that I have been doing independently on these same issues.

Cosmological effects on quantum objects

Briefly, I've been engaged in examining several different approaches based on ordinary quantum mechanics, which suggest that a quantum object at rest that is located within an expanding universe of finite duration will exhibit a finite uncertainty in location, and this uncertainty in location of the object will depend both on the expansion rate or age of the universe and on the mass of the object being localized.

A criterion for classicality and the quantum-classical transition

So, what does localization have to do with classical behavior? And how can we set up a criterion for classicality?

A classical object has a well-defined position in space as well as a well-defined momentum, whereas quantum objects exhibit uncertainties in either or both of these parameters. We are going to simplify by limiting our attention to objects at rest. For a quantum object at rest to behave classically, it must become localized. The extent of localization gives us a handle on the degree to which an object behaves classically. If we can introduce a useful parameter to characterize the extent of localization of an object, it can give also us a measure of to what degree an object behaves in a classical manner.

This appears to be easiest to accomplish for ordinary extended objects, which can be characterized at least roughly by their actual physical extent or size in space. We simply compare the size of the object with the size of the region of quantum uncertainty. If the region of quantum uncertainty is far larger than the size of object, and extends way beyond the edges of the object, then the overall behavior would resemble quantum behavior. On the other hand, if the region of quantum uncertainty for the location of the center of mass of the object is much smaller than the object's size, then we can expect that the object will behave in a more classical manner. So, when the size of the extended object is just equal to the size of the region of uncertainty of the center of mass of the object, this gives us a criterion for a threshold for classicality. There may be better ways to set a threshold marking a separation between quantum and classical behavior, but this is an easy one.

Evaluating the extent of localization of quantum objects

So, how can we evaluate the extent of spatial uncertainty, or the size of the localized region of high probability density for a quantum object that is confined within an expanding universe?

The phenomenon of localization of quantum objects in a universe that has been expanding for a limited duration of time has shown up in a number of quite different but straightforward studies of how quantum objects may behave in a temporo-spatially limited, expanding universe. Evidence for localization effects has been found by examining the effects of cosmological limitations using four quite different approaches:

- *Using Heisenberg uncertainty relations:* We consider the spread of Hubble expansion velocities within an extended object, and use the Heisenberg uncertainty relation to evaluate an associated spread in spatial uncertainty;
- *Using quantum wave packet behavior:* (Different approaches) An initially minimal Gaussian wave packet representing a quantum object at the time of the creation of the universe will disperse over the Hubble time to have a contemporary spatial width. Or, if a wave packet in the contemporary universe is formed from monoenergetic quantum wave functions that are truncated in time by the finite duration of the universe, it will exhibit an associated spatial width that is similar in size;
- *Using the Schrödinger Equation:* Starting with the Schrödinger equation and introducing in place of the difference between total energy and potential energy a kinetic energy term based on Hubble expansion velocities, we find wave function solutions that turn out to be spatially localized radial functions;
- *Using stochastic quantum mechanics:* We can take the point of view of stochastic quantum mechanics, and examine the size of the diffusion region corresponding to the Brownian type motion becomes over the lifetime of the universe, which leads to a measure of the present size of the region of localization describing the probability distribution of the quantum object.

All of these approaches seem to lead to roughly the same result: In an expanding, spatio-temporally limited universe, a quantum object will exhibit an uncertainty in location, or a quantum wave function will exhibit a pronounced concentration of its probability density, and the values for the sizes of these regions of uncertainty in location or regions of high probability density, all come out roughly the same.

Specifically, if the expansion and age of the universe are characterized by a Hubble constant $H_0$, a quantum object of mass m will be localized within a region of space with a linear size that is given approximately by the quantity $(h/mH_0)^{1/2}$, where h is Planck's constant. This can also be expressed in terms of the Hubble time $T_0$ which is the inverse of the Hubble constant and is approximately equal to the age of the universe, which gives a perhaps more intuitive expression, as the quantity $(hT_0/m)^{1/2}$. Thus, all of these studies indicate that a quantum object will be localized by cosmological effects, and that the approximate linear size of the region of localization depends on both the age of the universe and the mass of the object, with the size of this region of localization being given roughly by the quantity $(hT_0/m)^{1/2}$.

Discussion

Does this make any sense? I think so. If cosmological effects are causing obligatory classical behavior of objects in our world, then all objects above the threshold would always have to behave classically. If you put in the numbers, the threshold size that is obtained from these calculations for ordinary objects turns out to be about 0.1 millimeter, which for objects of ordinary densities corresponds to a threshold mass of about a microgram. That result seems to fit our common experience fairly well, as all objects above that size do seem to behave largely classically as entire objects. So these results are indeed telling us that all objects that we perceive in our human-sized world, even if they are fundamentally quantum mechanical objects, must appear to behave classically as a result of cosmological effects.

We know of course that even smaller objects can under some circumstances behave classically due to decoherence effects and other effects that seem to be able to bring about what amounts to classical behavior. So even smaller objects can and under some circumstances do behave classically. But the cosmological effects seem to set a threshold limit above which all objects must and do behave classically. (It should be noted that although Davies obtains qualitatively similar results, he finds that classicality should set in at considerably smaller sizes, - corresponding to a threshold object composed of about 400 quantum particles.)

But what about possible objections, such as the existence of quantum correlations between entangled quantum objects that can be located at great distances from each other – are these ruled out by the results above? By no means. This size threshold for classicality is informative mainly for the case of compact extended objects, but there appear to be no inconsistencies between the existence of this size threshold associated with classicality and the presence of quantum correlations between distant entangled quantum objects. That is because the constraint of the cosmological threshold is based on the uncertainty of localization of the center of mass of an object, or in this case, a system of objects. Thus, if we start with an original quantum object that subsequently separates into two entangled objects, the two entangled objects can move indefinitely far from away from each other and still behave as quantum objects, and the results of these calculations would just be providing us with a measure of what amounts to the uncertainty in the location in the center of mass of the system, not of the separation of the objects. There is no size constraint on how far away from each other such entangled quantum objects can move without being required to exhibit classical behavior.

I'll conclude by emphasizing that these results would appear to be of importance because the predictions of classicality deriving from cosmological effects would be absolute – there is no getting around them by manipulating local environmental conditions, as can be done in affecting quantum to classical transitions caused by local decoherence effects. Thus, these cosmological effects would appear to be fundamental, constituting absolute limitations, upper limits on quantum behavior in our universe.

…

ADDENDUM

To give you a brief description of one of the studies: Cosmological effects on a quantum system are explored by considering an object at rest in space with a universal Hubble expansion taking place away from it. Starting with the time-independent Schrödinger equation, we introduce in place of the difference between total energy and potential energy a kinetic energy term based on the radial velocity corresponding to Hubble expansion, and thus develop a governing differential equation which incorporates an intrinsic speed of expansion dependent on radial distance. Solving this governing equation leads to wave functions which turn out to Bessel functions of fractional order that exhibit pronounced central localization; these oscillatory radial wave functions are large near the origin of coordinates and drop off appreciably at distances comparable to the quantity $(h/mH_0)^{1/2}$. The size of the region of high probability density thus depends on both the Hubble constant and the mass of the object; objects with small masses tend to behave in a delocalized manner as ordinary quantum objects do in a static space, while objects with large masses have quantum wave functions that are concentrated into much smaller regions. And this result is in agreement with the other studies that I just mentioned that examine this question from the other points of view.